\documentclass[twocolumn]{aastex701}

\usepackage{xspace}
\usepackage{multirow}

\newcommand{\asec} {$^{\prime\prime}$}

\newcommand{\axaf}{\mbox{\em Chandra\/}}

\newcommand{\rubin}{\mbox{\em Rubin\/}}

\newcommand{\ska}{\mbox{\em SKA\/}}
\newcommand{\euclid}{\mbox{\em Euclid\/}}
\newcommand{\asdf}{\mbox{\em Roman\/}}
\newcommand{\mglens}{MG~B2016+112\xspace}
\newcommand{\marx}{\mbox{\textsc{Marx} \/}}
\newcommand{\saotrace}{\mbox{\textsc{SAOTrace} \/}}
\usepackage{graphicx}
\usepackage{hyperref}
\usepackage{color}

\usepackage{lineno}
\modulolinenumbers[1]
\begin{document}

%% Reintroduced the \received and \accepted commands from AASTeX v5.2
\received{}
\revised{}
\accepted{\today}

%% 
%%%%%%%%%%%%%%%%%%%%%%%%%%%%%%%%%%%%%%%%%%%%%%%%%%%%%%%%%%%%%%%%%%%%%%%%%%%%%%%%

\shorttitle{\axaf\ X-ray Spectra of the Lensed AGN's \mglens}
\shortauthors{D. A. Schwartz, J. Sisk-Reyn\'{e}s, \& A. Barnacka}

%%%%%%%%%%%%%%%%%%%%%%%%%%%%%%%%%%%%%%%%%%%%%%%%%%%%%%%%%%%%%%%%%%%%%%%%%%%%%%%%
\graphicspath{{./}{figures/}}
%% This is the end of the preamble.  Indicate the beginning of the
%% manuscript itself with \begin{document}.

%\begin{document}

\title{\axaf\ X-ray spectral evidence of the dual AGN nature of the gravitational lens \mglens at $z=3.273$ with 175 pc separation.}

\author[0000-0001-8252-4753]{Daniel A. Schwartz}
\affiliation{Center for Astrophysics $\vert$ Harvard \& Smithsonian, 60 Garden St, Cambridge, MA 02138, USA}
\email{dschwartz@cfa.harvard.edu}

\correspondingauthor{Daniel A. Schwartz}
\email{das@cfa.harvard.edu}

\author[0000-0003-3814-6796]{J\'ulia M.~Sisk-Reyn\'{e}s}
\affiliation{Center for Astrophysics $\vert$ Harvard \& Smithsonian, 60 Garden St, Cambridge, MA 02138, USA}
\affiliation{Department of Physics, University of Maryland Baltimore County, 1000 Hilltop Cir, Baltimore, MD 21250, USA}  
\email{julia.sisk_reynes@cfa.harvard.edu}

\author[0000-0001-5655-4158]{Anna Barnacka}
\affiliation{Center for Astrophysics $\vert$ Harvard \& Smithsonian, 60 Garden St, Cambridge, MA 02138, USA}
\email{abarnacka@cfa.harvard.edu}

\begin{abstract}

We use a new 175 ks \axaf\ X-ray observation of the $z=3.273$ radio-loud gravitational lens system \mglens\ to present definitive X-ray spectral evidence that it contains a  dual active galactic nucleus (AGN). With a projected
separation of 175 parsecs, this is the tightest separation confirmed AGN pair at high redshift. Prior multi-epoch very long baseline interferometry (VLBI) and archival \axaf\ data had revealed two distinct emitters, but could not rule out a single AGN with a complex, highly magnified jet. Our new \axaf\ observations allow a separate spectral analysis of each source. We find strong intrinsic absorption ($N_\mathrm{H} > 4\times 10^{22} \mathrm{cm}^{-2}$, 3$\sigma$ confidence) at the source redshift for the highly magnified source inside the caustic. 
This identifies it as an obscured AGN. 
In addition to the previously known type II AGN outside the caustic, this confirms a compact dual AGN system in \mglens. On-going and future surveys using facilities such as  \euclid, \ska, and the \rubin\ and \asdf\ observatories will uncover more such systems, probing SMBH evolution and gravitational wave precursors.
\end{abstract}


\section{Introduction} \label{sec:intro}

Hierarchical models of galaxy formation posit that major galaxy mergers drive the assembly and activation of supermassive black holes (SMBHs). Mergers can fuel  active galactic nuclei (AGN) and regulate star formation, activity that  peaked during cosmic noon \citep[$z\sim2-3$;][]{Hopkins2008,DiMatteo2008}. In major mergers each galaxy is expected to have a SMBH, so that the merged galaxy will contain a pair of black holes. To identify such pairs, both must be in a luminous active state. Direct evidence for dual AGN at high redshifts remains scarce, with confirmed pairs limited to separations of several kiloparsecs or more \citep{Chen2022,mannucci2022_gmp,mannucci2023_gmp,Li2025,Perna2025}. This observational gap stems from angular resolution limits: even the Hubble Space Telescope (HST) struggles below $\sim50$ milliarcseconds (mas)  \citep{NASAHST2023}, corresponding to $\sim400$ pc at $z=3$, rendering sub-kpc pairs difficult to discover via direct imaging.

Gravitational lensing mitigates this spatial resolution limit,  acting as a telescope to boost spatial resolution and to reveal otherwise inaccessible structures \citep{Barnacka2018,Barnacka2017}. For lensing configurations with a  source within and near to the inner caustic of a lens, magnifications can exceed hundreds, spatially amplifying compact features such as 10 -- 1000 pc scale jets and outflows, and revealing dual AGN. 

The radio-loud source MG B2016+112\footnote{We will use the discovery name \mglens in this paper. In J2000 coordinates the system  would be centered at J201918.1+112713.} at $z=3.273$ exemplifies this potential: primarily lensed by a foreground galaxy at $z=1.01$ \citep{Schneider1986}, it features a fold caustic that quadruply images a source inside and near to the caustic with magnification of a factor  $\mu\sim350$,  while doubly imaging another source that is outside the caustic \citep{Spingola2019}. We will refer to these as ``source 1'', and ``source 2'', respectively, following the designation of  \citet{schwartz2021}.

The presence of an AGN in the system was historically determined using both optical and radio observations \citet{Schneider1986}. Subsequent
optical spectroscopy one of the images of source 2 showed emission lines broadened by 500-900 km s$^{-1}$ indicative of photo-ionization complexity, and revealing source 2 is a type II AGN \citep{Yamada2001,Koopmans2002b}. \citet{Yamada2001} also detected  Ly-alpha, CIII], HeII], CIV] and NV] in image C. However, \cite{Yamada2001} interpreted MG B2016+112 as being a single AGN, with image C being the narrow-line emission region of this single AGN.  Later, source-plane reconstruction using multi-epoch VLBI observations favored the interpretation that this multiply-imaged system is comprised of two jetted radio sources (each with a flat- and a steep-spectrum component) separated by a projected distance of 175 pc at the source redshift ($0\farcs022$) \citep{Spingola2019}. \citet{Spingola2019} measured proper motions perpendicular to each other for the two radio sources, and different spectral indices, strengthening the dual AGN hypothesis. However, the interpretation of a single AGN outside the caustic (source 2; model-predicted magnification $\mu\sim 2$) with a jet crossing the caustic and producing highly magnified knots mimicking the second core (that of source 1; magnification $\mu\sim350$) could not be ruled out, as suggested by \citet{Koopmans2002b} and discussed by \citet{Spingola2019}.

Astrometric analysis exploiting non-linear angular amplification by gravitational lensing and using an archival 7.8 ks \axaf\ observation confirmed two separate X-ray emission regions consistent with the two VLBI sources \citep{schwartz2021}. However, with only 24 photons, the spectral analysis of \cite{schwartz2021} lacked statistical power to measure  intrinsic absorption from surrounding gas in the vicinity of the SMBH, thus distinguishing an AGN from jet knots \citep[which are typically unabsorbed;][]{Massaro2011}. 

Since source 2 was already known to be an AGN, the key to the investigation was achieving sufficient statistical power to detect absorption toward source 1, whose AGN nature had not previously been confirmed. Here, we present a new 175 ks \axaf\ dataset, revealing  spectra that resolve this long-standing debate and anchor models of high-$z$ SMBH mergers.

We use $H_0 = 67.4\ \mathrm{km\ s^{-1}\ Mpc^{-1}}$, $\Omega_\mathrm{m} = 0.315$, and $\Omega_\Lambda = 0.685$ \citep{Planck2020} throughout our analysis. 

\section{\axaf\  observations}\label{J2019}
To address the question of whether source 1 presents intrinsic X-ray absorption at sufficient statistical significance (which would rule out its radio jet interpretation and subsequently confirm its AGN nature), we obtained a 175 ks (165 ks live time)  \axaf\ observation ($\sim$22$\times$ longer than the archival). The data were obtained via 10 separate pointings over a fifteen-month period from December 2023 through March 2025  (\dataset[doi:10.25574/cdc.436]{https://doi.org/10.25574/cdc.436}).  Table~\ref{table:chandra_observations} lists parameters of the 10 new \axaf\ observations.

\begin{figure}[t]
\centering
 \includegraphics[width=3.7in]{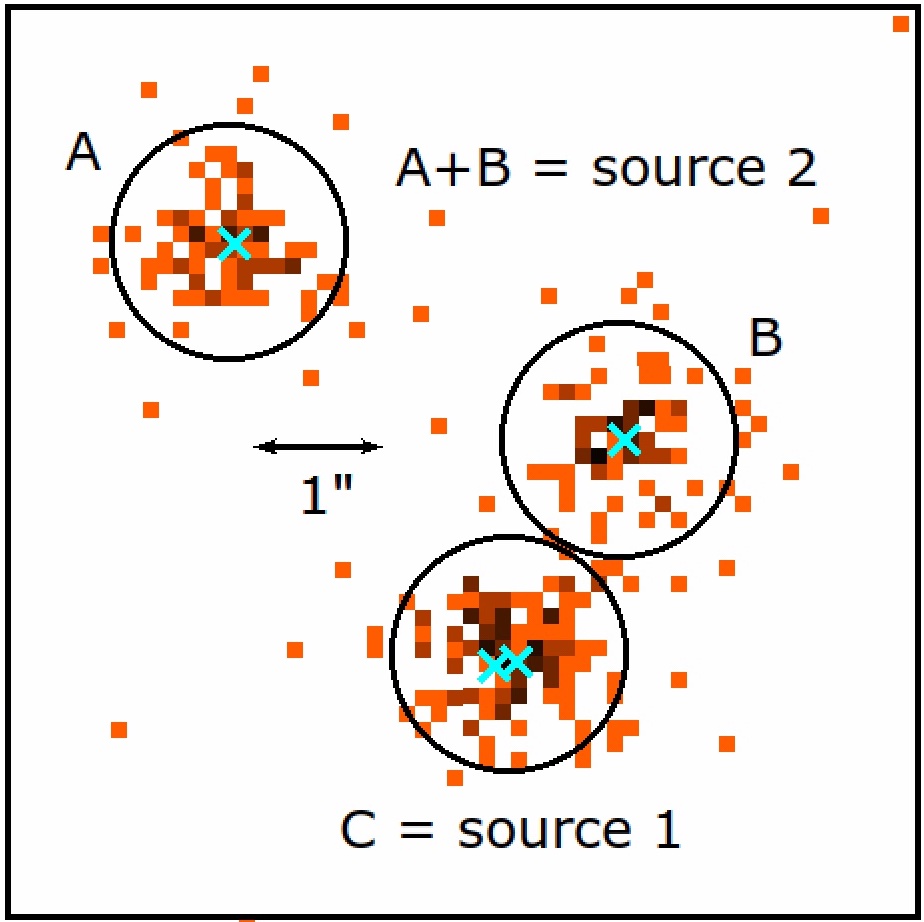}
\caption{The image plane merged data of the 165 ks live time observation of \mglens. The box shows a 7\asec $\times$ 7\asec region around the lensed images. The circles around each image have an 0\farcs90 radius, approximately the 80\% encircled counts radius over the 0.5 -- 7 keV band, as determined by an SAOTrace \citep{Jerius2004a} plus Marx \citep{Davis2012}~simulation of the individual measured image spectra (A--C) over this band. The ACIS-S3 0.5 -- 7 keV data is binned into 1/4 CCD pixels, 0\farcs123 on a side, where orange (black) pixels show 1 (6) counts per bin. White space has zero counts. Background is approximately 0.0025 counts per 1/4 pixel bin. The cyan crosses show where the \axaf\ aspect solution gives the ICRS coordinates of the VLBI components \citep[cf.][]{Spingola2019}. These offsets from the X-ray centroids are consistent with the absolute celestial location accuracy of \axaf.  \label{fig:J2019imageRevA}}
\end{figure}

At the half-arcsecond angular resolution of \axaf\ this system consists of three well resolved images, with source 2 lensed into images A and B separated by $\sim$3\farcs4, and source 1 lensed into two images that are unresolved in image C (overlapping cyan crosses in Figure \ref{fig:J2019imageRevA}), about $\sim$2\farcs0 from the X-ray image B in Figure~\ref{fig:J2019imageRevA}. The remaining two images of source 1 are very near A and B, but too faint to resolve. 

The first 7.8 ks \axaf\ observation of \mglens was  used to estimate a power-law spectrum with intrinsic absorption of 3 -- 85 $\times$ 10$^{22}$ H-atom cm$^{-2}$ at the source redshift when the power law index was fixed at $\Gamma$=2.0 for the combined spectrum of the  A + B (source 2)  images \citep{Chartas2001}.  The spectral re-analysis of \citet{schwartz2021} used 24 X-ray photons adding the three images and, assuming a power law photon index of 1.7, confirmed intrinsic absorption of $\sim$10$^{23}$ H-atom cm$^{-2}$ for the lumped spectrum.

\section{\axaf\  spectra}\label{J2019observations}

We used \texttt{ds9} \citep{Joye2003} to display each of the ten new observations and placed an 0\farcs90 radius circle about each of the distinct lensed images A, B, and C, where the blended X-ray image C corresponds to two radio images identified by \citet{More2009}. These circles have about an 80\% encircled-count-fraction radius. We used the \texttt{ciao 4.17} \citep{Fruscione2006} tool \texttt{dmhedit} to correct the RA\_NOM and DEC\_NOM values of each observation by the difference between the VLBI coordinates of the images and the celestial coordinate that the \axaf\ aspect solution reported for the centroid\footnote{We determine the centroid using the center of mass algorithm in ds9 after drawing a 0\farcs9 radius circle by eye around each image. We set 50 iterations, and tolerance of 0.1 pixel per iteration.} of each circle. More precisely, the average difference for the three images was used. We used the script \texttt{reproject\_events }to correct the tangent plane of each obsid to that of Obsid 28556, and then \texttt{dmmerge} to merge all 10 observations. ObsID 28556 was chosen as the reference based on having the longest exposure time.  Figure~\ref{fig:J2019imageRevA} shows the merged image with the regions selected for spectral analysis. 

Only \axaf\ has sufficient angular resolution to spatially resolve different lensed images of gravitationally lensed AGN systems (with image separations $\lesssim 1\farcs0$). For \mglens we carry out separate spectral analyses for the spectra of source 1 and of source 2. 

\begin{table}
\begin{center}
\caption{\textit{Chandra} X-ray  observations of \mglens. \label{table:chandra_observations}}
\begin{tabular}{|c|r|r|c|c|}
    \hline
    Obs ID&Live time (ks) & Counts\tablenotemark{a} & Start Date& Group\tablenotemark{b} \\
    \hline
 %   429&7.8&24&2000-04-12 &    \\
   28554&9.54 & 17 & 2023-12-22&  1  \\
    29135&9.95 & 14 & 2023-12-23&  1  \\
    29136&9.95 & 16 & 2023-12-23&   1 \\
    28556&29.68 & 49 & 2024-10-06&   2 \\
    28141&25.73 & 46 & 2024-10-15&    2\\
    30569&25.73 & 45 &  2024-10-16&    2\\
    28555&14.40 & 36 & 2024-12-04&  3\\
    30645&14.88 & 34 & 2024-12-04&   3 \\
    28557&10.93 & 18 & 2025-03-10&    4\\
    30844&13.89 & 31 & 2025-03-11&    4\\
    \hline
    Sum&164.68 & 306 & & \\
    \hline
\end{tabular}
\end{center}
\tablenotetext{a}{Number of X-ray counts (0.5 - 7 keV observed band) summed in circles of radii 0\farcs90 about all three images in each ObsID.}
\tablenotetext{b}{The observations within one month are grouped for the light curve displayed in Figure~\ref{fig:J2019timeHistory}.} 
\end{table}

\begin{figure}[t]
\centering
 \includegraphics[width=3.2in]{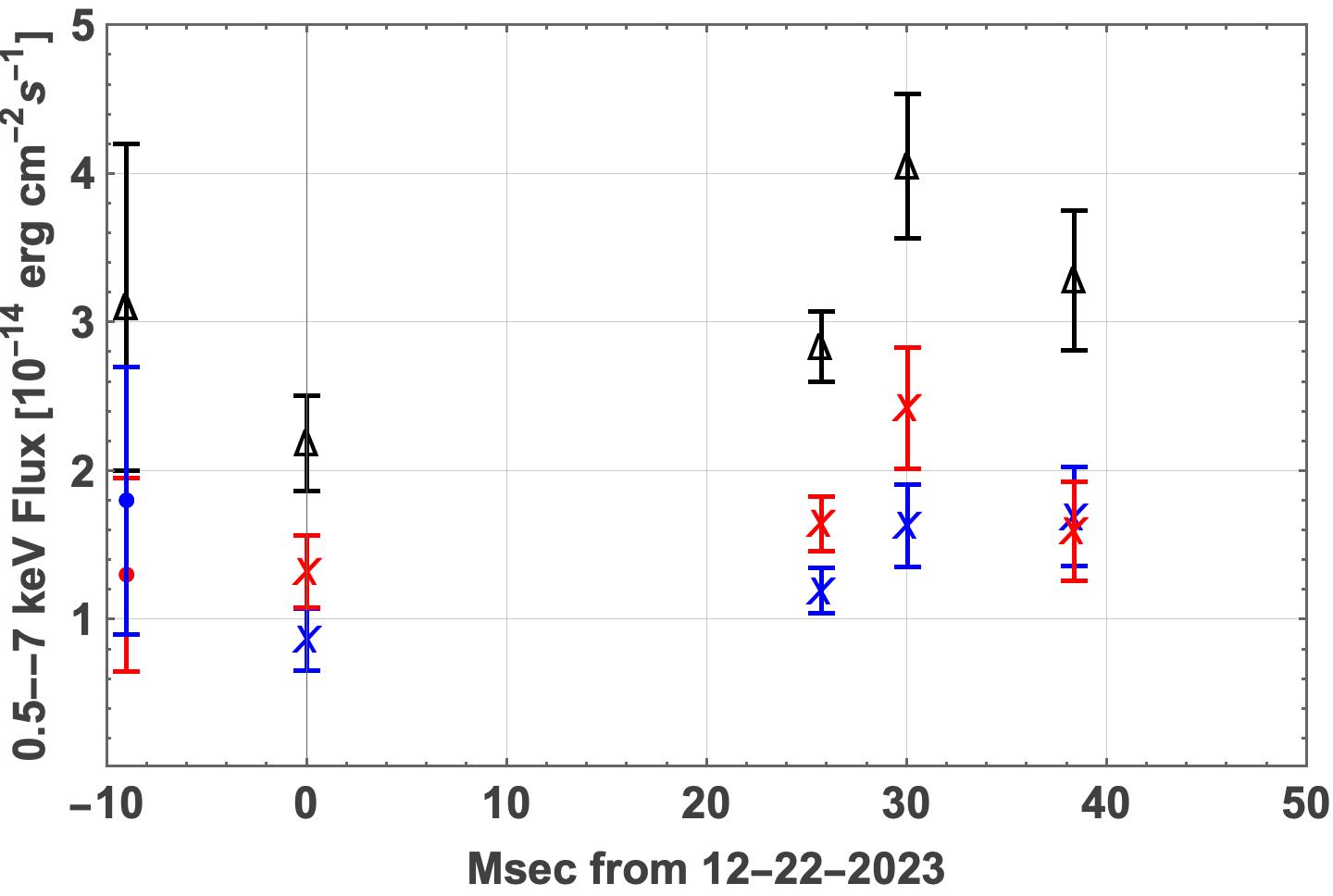}
\caption{Light curve of all \axaf\ observations of each of the two X-ray sources of \mglens. Observations are grouped as indicated in column 5 of Table~\ref{table:chandra_observations}. Blue "X" points are source 1, image C. Red "X"  points are source 2, the sum of images A and B.  Error bars are 1$\sigma$ derived from the square root of the number of counts. We plot the measured 0.5 -- 7 keV flux, i.e., the magnified flux from 2.1 -- 30 keV in the quasar rest frame,  in units of 10$^{-14}$ erg cm$^{-2}$ s$^{-1}$. Time is measured positively from 23 December 2023. The points plotted at -9 Ms were measured in the archival 7.8 ks observation on 12 April 2000. The black triangles sum the measured counts from all images, as is necessary to compare to observations by any other X-ray telescope. \label{fig:J2019timeHistory}}
\end{figure}

Figure~\ref{fig:J2019timeHistory} shows the light curve of the measured (magnified) 0.5 -- 7 keV flux of the two sources, grouped as identified in Table~\ref{table:chandra_observations}. The flux is derived by fitting each observation to a power-law plus absorption by our galaxy, plus intrinsic absorption at the source redshift. As described below, this is the best fit for each source, although the data does not rule out that source 2 has no intrinsic absorption.  To convert to intrinsic luminosity, source 1 (2) -- blue (red) in Fig \ref{fig:J2019timeHistory} -- should be divided by $\sim$350 ($\sim$2) to correct for the lensing magnification \citep{Spingola2019}, giving intrinsic rest frame 2~ --~ 30 keV luminosities of 3.6$\times 10^{42}$ and 8.3$\times 10^{44}$ erg s$^{-1}$ for source 1 and 2, respectively. Neither source is significantly variable, and the two source fluxes are consistent with being the same in each observation.  The near equality of the two measured fluxes must be a coincidence, since their magnification factors are so different. \emph{XMM-Newton} observed this source three times in October 2002.
Converting the 0.2-12 keV absorbed fluxes listed on the ESA/XMM archive to the 0.5 -- 7 keV band using a fixed $\Gamma=1.7$ and intrisic absorption n$_{H}$=3.4$\times 10^{23}$cm$^{-2}$ gives an average flux $(3.3~\pm~0.3)\times {10}^{-14}$ erg cm~$^{-2}$~s$^{-1}$, consistent with the flux measured by \axaf\ being constant over the entire $\sim25$-year observation interval (5.7 years in the source frame; Figure \ref{fig:J2019timeHistory}).

We used the \texttt{ciao 4.18} tool \texttt{specextract} to recover photons in regions of  0\farcs9 radius about each of the three images, for each observation interval (Table~\ref{table:chandra_observations}). The tool formulates the correct absolute response function (the area response file or ARF) and redistribution matrix (RMF) for the appropriate observation date and the actual location on the ACIS-S CCD. We used \texttt{combine\_spectra} to create a single spectral file for each source.  This yielded 139 and 167 total counts for source 1 and source 2, respectively. The background is 0.16 count arcsec$^{-2}$, so only 1 count out of the 306 used for spectral analysis.   Each source has its corresponding files extracted from each ObsID, and each spectrum is analyzed jointly for the 10 observation intervals. 

We used \texttt{sherpa 4.18.0} \citep{Siemiginowska2024} and \texttt{CALDB 4.12.3} to do the spectral fitting, and report the one sigma error on parameters unless otherwise noted. We use the ``Cash'' statistic on unbinned data \citep{cash1979}. All fits used a Galactic absorption frozen at  n$_H$ = 1.3$\times$10$^{21}$ H-atom cm$^{-2}$ \citep{HI4PI}. Table~\ref{table:spectra} shows all of our spectral fitting results.

%\begin{center}
\begin{deluxetable*}{llrclr}
\caption{Fits to spectral models of the two \mglens sources. \label{table:spectra} }
 %   \multicolumn{3}{|c|}  
 %  & Source 1 (C)& Source 2 (A+B)\\
 \tablehead{ \multicolumn{1}{c}{}&\multicolumn{2}{c}{Source 1 (C)}&\multicolumn{1}{c}{}&\multicolumn{2}{c}{Source 2 (A+B)} \\
 \cline{2-3}
 \cline{5-6}
   \vspace{-0.1in}
  &\colhead{photon}& & & \colhead{photon}& \\
 \colhead{model} & \colhead{index $\Gamma$} & \colhead{n$_H$\tablenotemark{a}}& &\colhead{index $\Gamma$} & \colhead{n$_H$\tablenotemark{a}} }
   % \hline
   \startdata
    Power Law (PL)&1.12$\pm$0.20& - & & 1.07$\pm$0.18& -  \\
    PL + Abs\tablenotemark{b}&2.54$\pm$0.21
    &72$\pm$20\tablenotemark{d} & &1.46$\pm$0.18
    &19$^{+15}_{-12}$\tablenotemark{d} \\
    PL + Abs\tablenotemark{b}&1.7\tablenotemark{c}&  34$\pm$10& & 1.7\tablenotemark{c}&30$^{+18}_{-10}$\tablenotemark{d}\\
  \enddata
  \tablenotetext{a}{Intrinsic absorption at the source redshift, in 10$^{22}$ H-atom cm$^{-2}$}
  \tablenotetext{b}{Power law plus intrinsic absorption. See Figure~\ref{fig:correlation} to best assess the errors on the correlated slope and absorption parameters.}
  \tablenotetext{c}{Photon index fixed at 1.7}
  \tablenotetext{d}{1$\sigma$ error assuming the photon index is fixed.}
\end{deluxetable*}
%\end{center}

We first fit a simple power law using the photon spectrum $dN/dE=A e^{-\sigma(E) * n_H} E^{-\Gamma}$; i.e., the model \texttt{xsphabs*powlaw1d} in \texttt{sherpa}.  %Figure~\ref{fig:index} shows the best fit power-law indices and error bars. 
The spectral indices for each observation or grouped observation of each source are consistent with the joint fit to all observations. 
Those photon indices for the joint fits to the ten observations are exceptionally flat: $\Gamma=1.12\pm0.20$ for source 1 and $\Gamma=1.07\pm0.18$ for source 2.

\begin{figure*}
\includegraphics[width=0.46\textwidth]{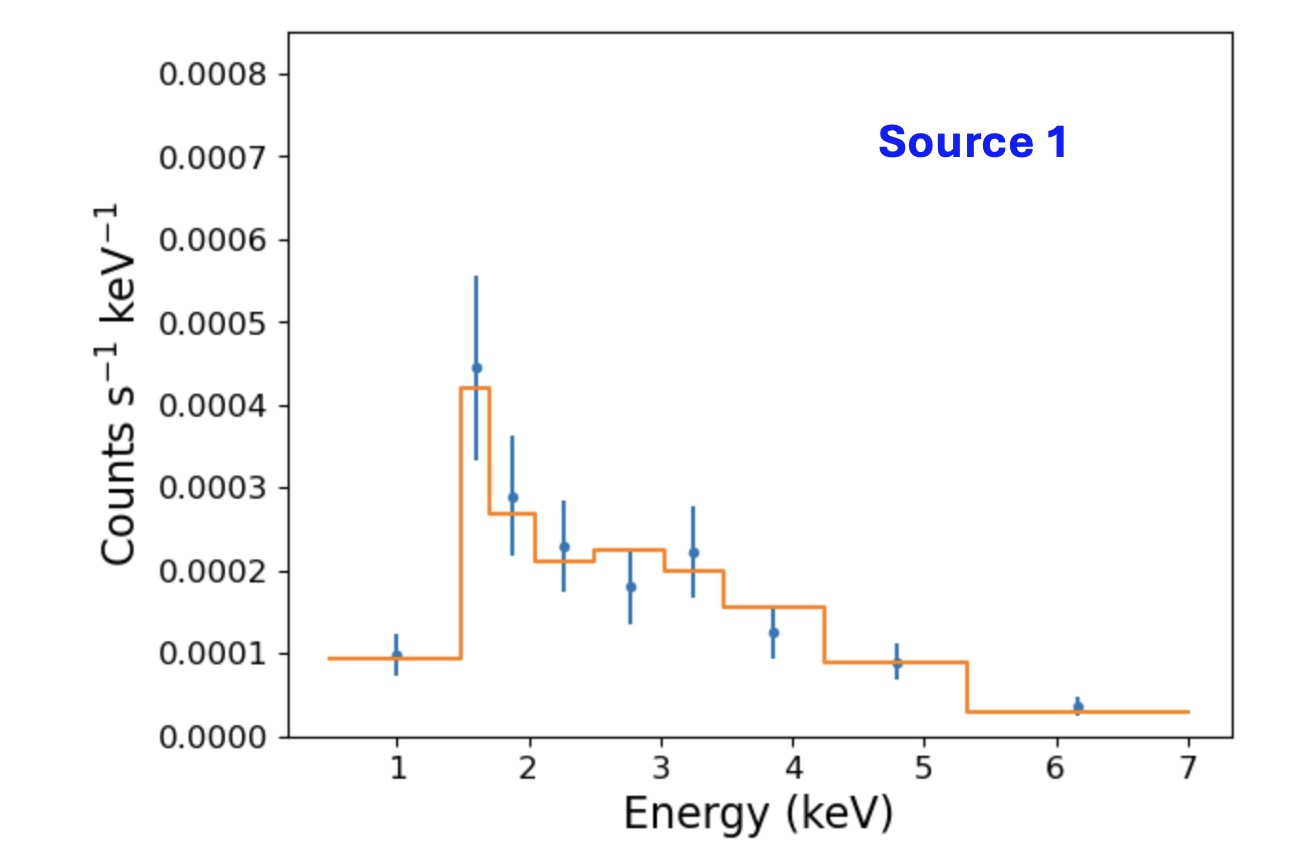}
\includegraphics[width=0.46\textwidth]{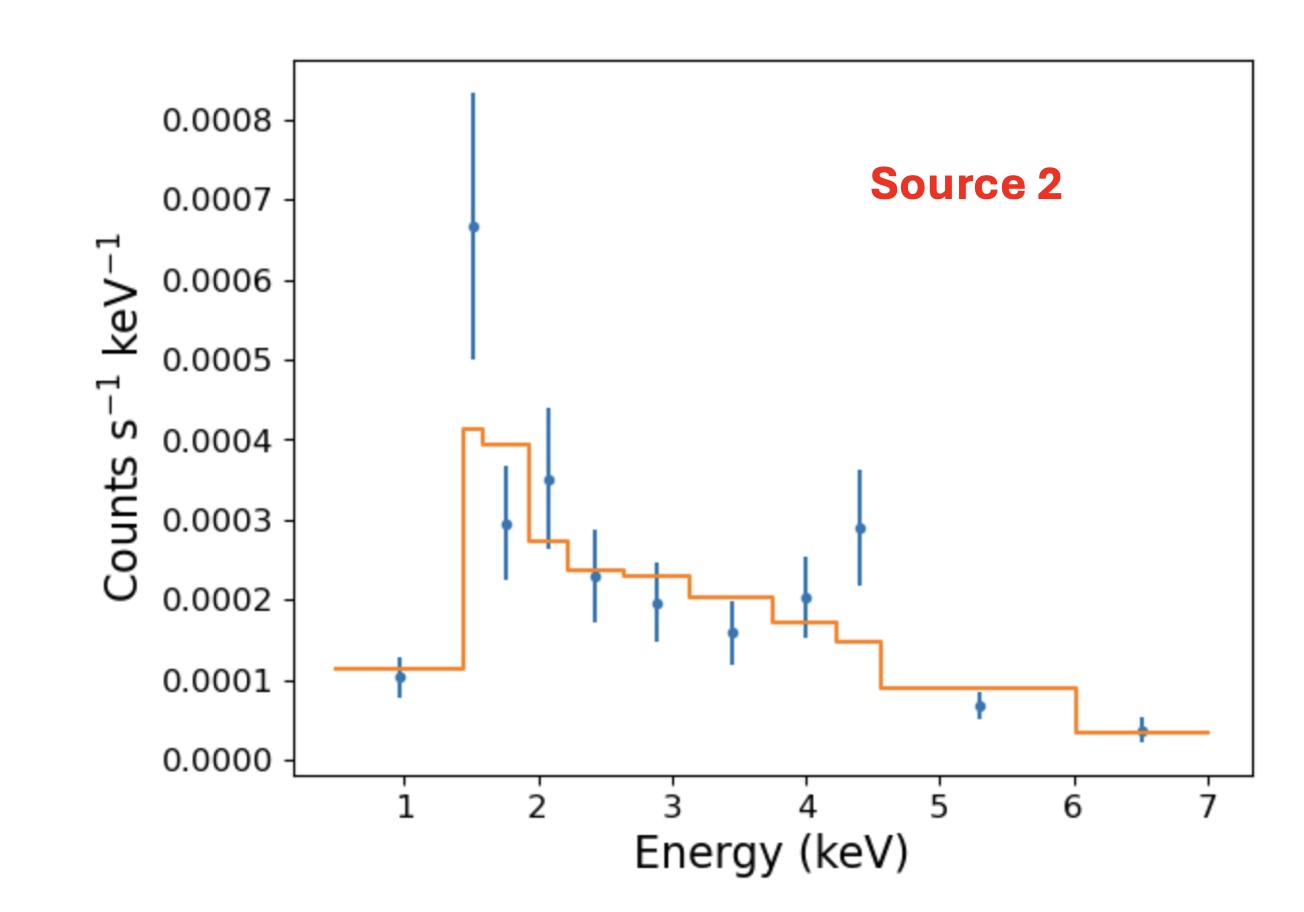}
\caption{The orange histograms show the models of a power-law with absorption in our galaxy, modified by intrinsic absorption local to the source at the redshift 3.273, folded through the expected mirror and detector response. The x-axis shows the observed energy. For the purpose of display only, the data have been binned into groups of 16 counts per energy channel. The error bars are 1$\sigma$ derived from the square root of the number of counts. The orange histogram shows the predicted counts per keV in each bin for the best fit photon power law index  $\Gamma$ and intrinsic absorption  n$_{H}$, and gives the width in energy for each point. Left panel: source 1 with  $\Gamma=2.54\pm0.21$ and n$_{H}$=72 $\times$10$^{22}$H-atom cm$^{-2}$. Right panel:  source 2 with $\Gamma=1.46\pm0.18$ and n$_{H}$=19 $\times$10$^{22}$H-atom cm$^{-2}$.
\label{fig:abs}}
\end{figure*}

The extremely flat spectrum is more likely due to intrinsic absorption associated with the source. We use the model\ \texttt{xsphabs*xszphabs*powlaw1d} fixing the redshift at z=3.273 and fitting each source separately for the intrinsic column density, photon index, and power law normalization. For  source 1 the photon index is $\Gamma=2.54\pm0.21$ with n$_{H}$=72 $\times$10$^{22}$H-atom cm$^{-2}$. The likelihood value relative to the fit with no intrinic absorption decreases by 10.5, so for the one degree of freedom added this fit is significantly preferred over a powerlaw with no intrinsic absorption to a confidence of 99.9\%, slightly higher than 3$\sigma$ probability. Source 2 is best fit with a photon index of $\Gamma=1.46\pm0.18$ and intrinsic absorption at the redshift of the source of n$_{H}$=22 $\times$10$^{22}$H-atom cm$^{-2}$. In this case the one additional parameter of intrinsic absorption only decreases the likelihood function by 1.3, which has a 25\% probability of happening by chance, so this fit is not a significant improvement over a simple power law. Figure~\ref{fig:abs} shows the spectra of the two sources fit with intrinsic absorption.

\begin{figure*}
\includegraphics[width=0.4\textwidth]{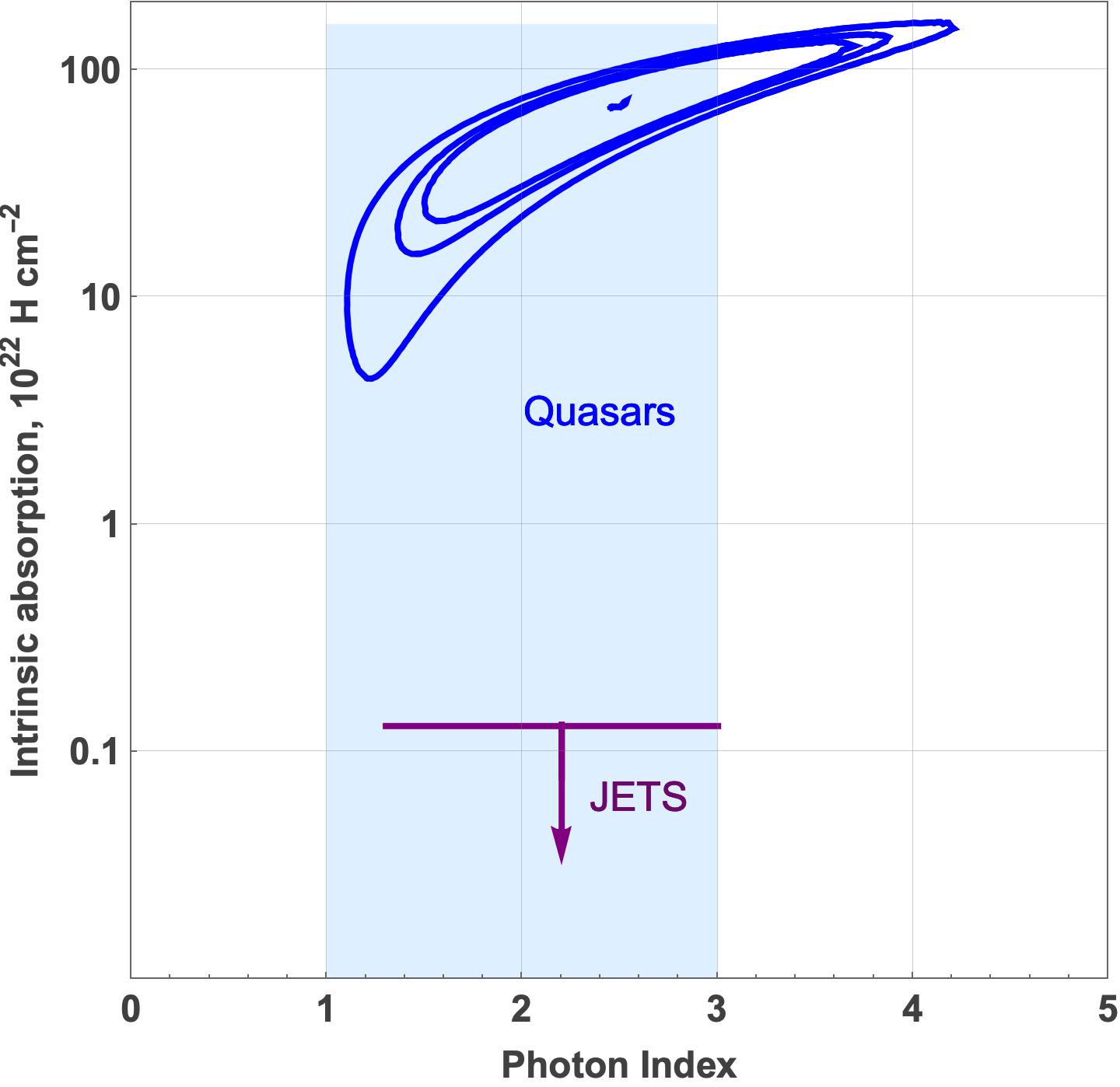}
\hspace{0.2\textwidth}
\includegraphics[width=0.4\textwidth]{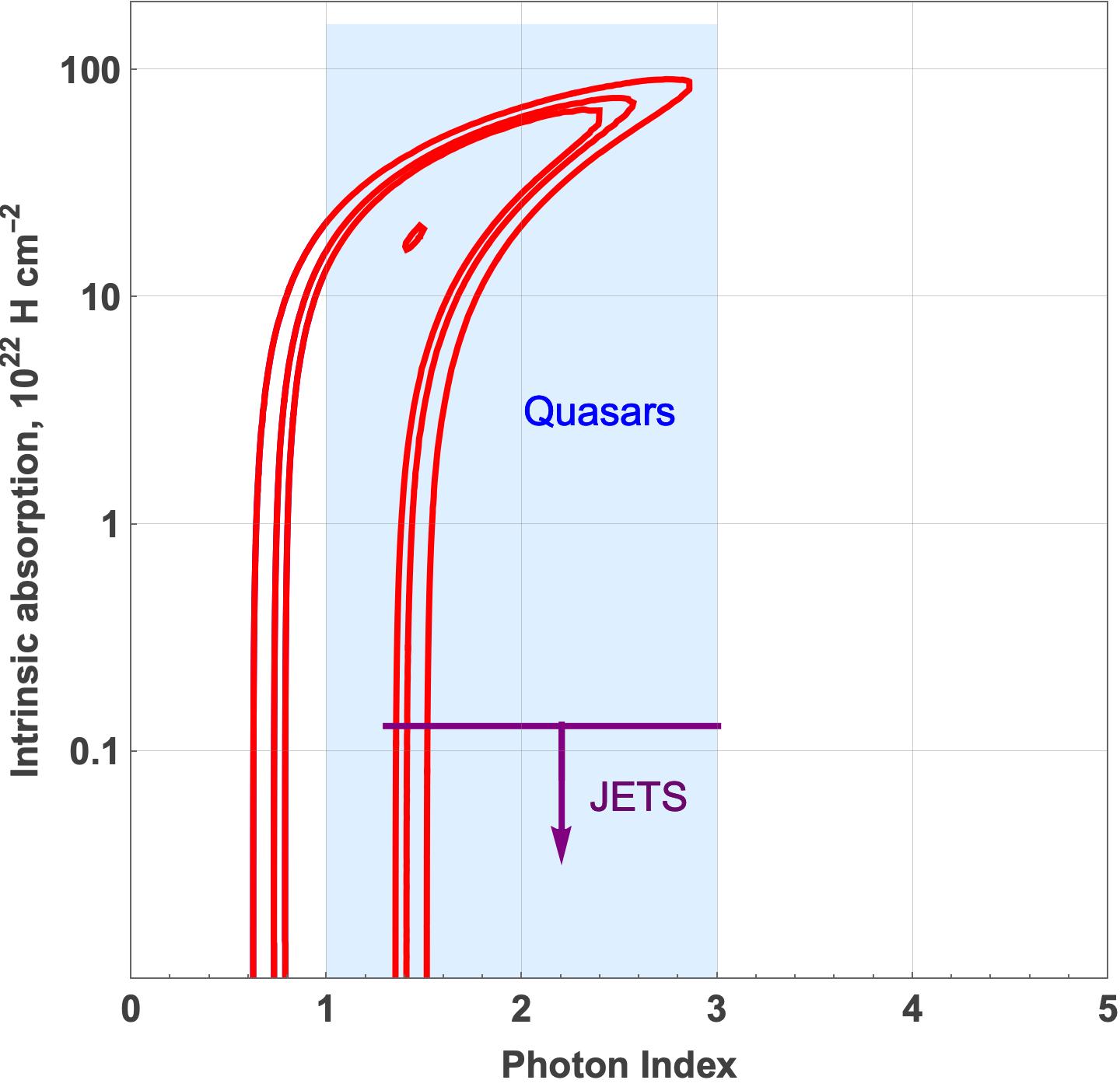}
\caption{The left panel shows source 1 and the right panel source 2. The contour plots show allowed combinations of the correlated parameters of intrinsic absorption and photon power law index.  The contours from innermost to outermost delineate regions of 1, 2, and 3 $\sigma$ probability that the data could arise from such a pair of parameters, when marginalized over the uninteresting parameters of photon index and the normalization. The central dots show the best parameters from the spectra in figure~\ref{fig:abs}. The shaded blue region is indicative of the location of most quasars. The purple down-arrow indicates the allowed region for jets, with the bar at the n$_{H}$=1.3$\times 10^{21}$ galactic column density that would allow no excess over the galactic column in this direction.  
\label{fig:correlation}}
\end{figure*}

We consider a maximum likelihood ratio test for the allowed values of the  n$_H$ parameter of interest. The $\Delta$Likelihood is distributed as $\chi^2$ with one degree of freedom \citep{Wilks1938}. The contours in Figure~\ref{fig:correlation} delineate the  regions that correspond to 1, 2, and 3 $\sigma$ probability of enclosing the true parameter region. Source 1 formally excludes n$_H$=0 to 3.2$\sigma$ equivalent probability, with a 3$\sigma$ lower limit n$_H > 4 \times 10^{22}$ H-atoms cm$^{-2}$. Source 2 is allowed to have zero absorption local to the source for a sufficiently flat photon index, $\Gamma < 1.7$, although within statistics its spectrum also could be the same as source 1.

 Quasars will generally be found in the shaded region, with indices between 1 and 3 \citep{2007ApJ...665.1004J,2009ApJS..183...17Y} and any amount of intrinsic absorption \citep{2002ApJ...571..234R,Risaliti2009,2017NatAs...1..679R}. Jet knots outside of the dusty torus should not show absorption substantially more than that in the line of sight through our galaxy \citep{2009A&ARv..17....1W}.  

%%%%%%%%%%%%%%%%%%%%%%%%%%%%%%%%%%%%%%%%%%%%%%%%%%%%%%%%%%%%%%%%%%%%

 \section{Discussion and conclusions}

The goal of this observation was to determine whether source 1 is an AGN.  
Source 2 was known to be an AGN due to the observation of broadened emission lines \citep{Yamada2001} and known to host a jet \citep{Spingola2019}. Multi-epoch VLBI lens modeling by \cite{Spingola2019} had preferred a dual AGN interpretation for sources 1 \& 2, but could not rule out that source 1 was a jet (emitted from source 2), being highly magnified ($\mu\sim 350$) when crossing the inner caustic.
The work of \citet{schwartz2021} confirmed the existence of two distinct X-ray sources, but the archival \axaf\ observations did not have sufficient counts to measure the individual X-ray spectra of each source. These results motivated the new \axaf\ observations we have presented, specifically to measure  intrinsic X-ray absorption in source 1. 

We measure a large intrinsic X-ray absorption of source 1, with a 3$\sigma$ lower limit $N_H > 4 \times 10^{22}$ H-atoms cm$^{-2}$, which unambiguously identifies source 1 as an obscured AGN. Absorption in AGN is generally attributed to a dusty torus, at distances less than 10 pc from the SMBH \citep[e.g.][]{Chen2023dust,Mandal2024}. The measured absorption cannot arise from reprocessing in the torus of source 2 (that is, at a distance of at least 175 pc in projection). Therefore, both X-ray emission regions identified by \cite{schwartz2021} originate from two separate AGN.

The ACIS-S response below 2 keV can be precluded as a source of potential bias in measuring the absorption. The ACIS-S response is increasingly reduced at energies below 2 keV, where intrinsic absorption effects are most pronounced. However, this is accounted for in the time-dependent calibration data base maintained by the \axaf\ X-ray Center \citep{Graessle2006}, and evident in the orange histograms of Figure~\ref{fig:abs} which show the predicted response to the given source models. Figures 6.4 and 6.8 of the relevant \axaf\ Proposer's Guide\footnote{https://cxc.harvard.edu/proposer/POG/arch\_pdfs/POG\_cyc27.pdf} show the ACIS-S3 response extending below 1 keV, and the good statistics from observation of the galaxy cluster Abell 1795 that are used to calibrate the response. We note that for recent ACIS observations of high redshift quasars both large intrinsic absorptions $>$ 10$^{22}$ cm$^{-2}$ are measured \citep[e.g.][]{Breiding2026} as well as measured limits less than 10$^{22}$ cm$^{-2}$ \citep[e.g.][]{Marlar2025} to column densities. Finally, we note that the present measurement of absorption is consistent with the one on 12 April, 2000 \citep{Chartas2001,schwartz2021} when the ACIS filter was essentially uncontaminated.

With the confirmation of source 1 as an AGN, we have confirmed the dual AGN nature of \mglens, making this the closest separation AGN pair known at high redshift. 
All massive galaxies are expected to harbor a central SMBH. When these galaxies merge, their SMBHs may experience a dual AGN phase. Dynamical friction may reduce the distance between the two SMBHs, to the extent that the two systems may eventually form a gravitationally bound pair and merge \citep{Li2020a,Li2020b}. The galaxy merging process is expected to provide gas supply, possibly triggering AGN activity in both SMBH \citep{burke-spolaor_2018_proc,2007Sci...316.1874M}.

Although cosmological simulations predict such systems \citep{2018MNRAS.475.4967T,2019MNRAS.483.2712R,volonteri2022,2023MNRAS.522.1895C}, direct observational evidence has remained elusive. Until now, dual AGNs with sub-kiloparsec separations had been observed only at low redshift, leaving scarce direct evidence for such systems at or beyond the peak of galaxy evolution \citep{2012ApJ...753...42C,2018MNRAS.480.1639D,Chen2022,2025ApJ...988..126C,2026arXiv260719491S}. The system \mglens presented here, with a projected separation of only 175 pc, is the first of its kind and provides a unique probe of hierarchical galaxy evolution. Its detection was enabled by strong gravitational lensing, which both magnified the source flux and spatially magnified the system. Without this combined flux and spatial magnification, the source would have been too faint and its two components too compact to resolve. Lensing therefore provided the sensitivity and effective source-plane resolution required to identify the two AGNs and measure their 175 pc projected separation at $z = 3.237$.

Determining the fraction of dual AGN and their circumnuclear environments is critical for estimating galaxy merger rates in the early universe, and also for assessing the time scales for SMBH pairs to coalesce and emit gravitational waves. \citet{derosa2019} and \citet{bogdanovic2022} review techniques for identifying dual and binary SMBH. Estimates of dual AGN found in a single galaxy subsequent to a merger are uncertain, with a range of at least 1\% \citep{volonteri2022} to 20 -- 30\% \citep{Perna2025}. 

Searches for dual AGN are currently of great interest, enhanced by the intial data releases from \euclid\ and \ska, and the commissioning of the \rubin\ and \asdf\ observatories. These facilities allow detections of pairs of radio and optical sources at separations below a kpc. \citet{Ulivi2025} applied machine learning methods to Euclid data to reveal quasar candidates that could be either dual AGN or lensed AGN. \citet{DAmato2026} make the case for the SKAO identifying pairs with projected separations of only several hundred pc at large redshift. 
One can then use high resolution X-ray imaging and spectroscopy to provide critical diagnostics for confirming AGN pair candidates as either dual or lensed systems. In parallel, gravitationally lensed AGN pairs at high redshift may be directly revealed via X-ray astrometry.

\vspace{5mm}

This work has been supported by NASA contract NAS8-03060 to SAO, NASA/ADAP grant 80NSSC24K0617, and grant GO4-25053X from the CXC. This research made use of the NASA Astrophysics Data System, now the SciX Digital Library, funded by NASA under Cooperative Agreement 80NSSC21M00561.  We thank D. Burke for assistance with \texttt{sherpa}. We thank Giulia Migliori, Cristiana Spingola, and Mauro Dadina for calling our attention to the XMM observation.
\vspace{5mm}
\facilities{Chandra (ACIS)}

\software{ciao-4.17, ciao-4.18 \citep{Fruscione2006}, sherpa-4.18 \citep{Siemiginowska2024,Burke2025},  SAOImageDS9 Version 8.4.1 \citep{Joye2003}, APLpy \citep{Robitaille2012}, \saotrace - 2.0.6.1 \citep{Jerius2004a}, \marx - 6.0.1 \citep{Davis2012}.
}

\bibliography{J2019references}{}
%\bibliography{references}{}
\bibliographystyle{aasjournal}
\end{document}